\documentclass[aps,prl,showpacs,twocolumn]{revtex4}
\usepackage{amsmath,amssymb,array}
\usepackage{graphicx}
\newcommand{\beq}{\begin{eqnarray}}
\newcommand{\eeq}{\end{eqnarray}}
\renewcommand{\vec}[1]{{\mathbf{#1}}}

\begin{document}

\title
{Bose Hubbard model in the presence of Ohmic dissipation}

\author{Denis Dalidovich and Malcolm P. Kennett }

\address
{Department of Physics, Simon Fraser University, 8888 University Drive, 
Burnaby, British Columbia V5A 1S6, Canada}

\date{\today}
\begin{abstract}
We study the zero temperature mean-field phase diagram of the 
Bose-Hubbard model in the presence of local coupling between 
the bosons and an external bath. We consider a coupling
that conserves the on-site occupation number, preserving the
robustness of the Mott and superfluid phases.
We show that the coupling to the bath renormalizes 
the chemical potential and the interaction between the bosons and reduces the 
size of the superfluid regions between the 
insulating lobes. For strong enough coupling, a finite value of hopping 
is required to obtain superfluidity around the degeneracy points where
Mott phases with different occupation numbers coexist. 
We discuss the role that such a bath coupling may play in experiments 
that probe the formation of the insulator-superfluid shell structure 
in systems of trapped atoms.
\end{abstract}


\maketitle

The Bose-Hubbard model (BHM) describes a wide variety of physical situations for
appropriate parameter ranges \cite{JJ}: Josephson junction arrays,
ultracold atoms in optical lattices \cite{BlochRev,Jaksch}, and polaritons
or photons in arrays of resonant optical cavities \cite{Polariton,Photon}.  A key feature
of the BHM is the existence of a transition between Mott
insulator and superfluid phases, recently observed for cold
atoms in optical lattices \cite{Greiner,Bloch,Gerbier3}.  This
 superfluid-insulator transition (SIT) is
well understood theoretically \cite{Fisher,Sachdev,Herbut} and
occurs as the magnitude of the inter-site hopping $J$ is varied relative
to the on-site repulsion $U$. At $T=0$, 
there are insulating Mott phases for small enough $J/U$ which exist in
well-defined domains (Mott lobes) in the $zJ/U$-$\mu/U$ 
plane ($\mu$ is the chemical potential and $z$ is the co-ordination), 
while in the large $J$ limit there is a superfluid 
state with fully delocalized particles. 

In some, or all, of the various systems that are described by the BHM, it is
likely that there will be coupling to external degrees of freedom in the
environment that can be modelled as a bath that lead to equilibration in the
BH system.
At small but finite temperatures, the insulating nature of the Mott 
phase does not change, although the on-site occupation numbers 
undergo thermal fluctuations leading to finite 
compressibility \cite{Dickerscheid}.
Experiments on optical lattice systems are in this regime, where temperature
is important \cite{Gerbier1}.  However, it is still a matter of debate
whether heating takes place as the depth of the lattice potential is
increased \cite{Gerbier2,Diener}. 
It has been suggested that the main cause for such a temperature rise
is the increase of the excitation energy gap combined with the necessity of
maintaining a constant entropy \cite{Gerbier1,Ho,Pollet}. 
If heating can be important, then maintaining the system at
a constant temperature may require the presence of some 
source of dissipation to remove energy.
The issue of dissipative dynamics in cold atom systems has attracted
considerable recent attention, in
the contexts of cooling atoms in the lowest Bloch band \cite{Griessner},
Bose-Fermi mixtures \cite{BoseFermi}, the decay of supercurrents 
\cite{McKay,Polkovnikov}, and one dimensional cold molecular
gases \cite{Syassen}.  
Hence, it is of great interest to address 
theoretically the issue of how coupling to a heat bath
affects the phase diagram of the BHM.

In this Letter we investigate the influence of coupling to a heat bath
(assumed to be Ohmic) on the standard phase diagram of the BHM
 \cite{Fisher,Sachdev,Herbut}. 
We assume that the system of bosons is in full equilibrium with the bath
at fixed temperature. Throughout our treatment, we 
assume for simplicity that $T=0$, but briefly mention
possible changes in the phase diagram due to finite temperature.
We consider a coupling between the bath and bosons 
through the particle operator $\hat{n}_i$, implying that the 
number of bosons in the system is fixed.
Hence the Mott insulator may still be defined as
a state with integer expectation value for the occupation number
$n_0$ on each site, and in the superfluid phase there is a finite 
phase stiffness.  The coupling to a bath leads to:
i) Renormalization of  $\mu$ and 
$U$ to $\mu^{\prime}$ and $U^{\prime}$  respectively, which 
determine $n_0$ in the Mott regions;
ii) the regions of superfluidity between the Mott lobes 
shrink in size on the $zJ/U$-$\mu/U$ phase diagram.
If the coupling to the bath is strong enough, superfluidity is absent 
for small $zJ/U$ regardless of $\mu$, illustrating that
coupling to external degrees of freedom inhibits the formation of a
phase coherent state.
We  discuss possible connections of our findings to measurements 
performed in the presence of a magnetic trap potential and suggest how 
the coupling to a bath might be detected.

The starting point of our calculations is the standard BH Hamiltonian,
and we assume that the bosons couple locally to the heat bath, so 
that the total Hamiltonian is 
\beq\label{hamilbos}
\hat{H} & = &\hat{H}_{bos}+ \hat{H}_{bath}+\hat{H}_{coup},
\eeq
where $\hat{H}_{coup}$ describes coupling to the bath.
The boson Hamiltonian $\hat{H}_{bos}$ consists of a local piece $H_0$ and a hopping term
$H_J$:
\beq\label{hamil1}
\hat{H}_{bos} &= & - J
\sum_{\langle i,j \rangle} (\hat{a}_i^\dagger \hat{a}_j 
+\hat{a}_j^\dagger \hat{a}_i)
-\mu \sum_i \hat{n}_i \nonumber\\ 
& & + \sum_i \frac{U}{2} \hat{n}_i \left( \hat{n}_i -1\right)= \hat{H}_J
+\hat{H}_0 , \\
\label{hamilbath}
\hat{H}_{bath} &= &\sum_{i\alpha} \varepsilon_\alpha \hat{b}_{i\alpha}^\dagger 
\hat{b}_{i\alpha},\\
\label{hamilin}
\hat{H}_{coup} &= &  \sum_{i\alpha} g_\alpha \left(
\hat{b}_{i\alpha}^\dagger + \hat{b}_{i\alpha} \right) \hat{n}_i.
\eeq
$\hat{n}_i= \hat{a}_i^\dagger \hat{a}_i$ is the on-site number
operator, and the notation $\langle i,j \rangle$ indicates that the
summation is restricted to nearest neighbors only.
The operators $\hat{b}_{i\alpha}^\dagger$ and  
$\hat{b}_{i\alpha}$ create and annihilate the bath degrees of freedom 
which have an à
energy spectrum $\varepsilon_\alpha$. 
This form of coupling, in which the bath degrees of 
freedom are coupled to the on-site density $\hat{n}_i$,
is most suitable to take into account fluctuations of 
the chemical potential. $g_\alpha$ characterizes the strength of coupling,
which we assume to be purely local and Ohmic:
\beq\label{inter}
g_{\alpha}^{2} = \eta \varepsilon_{\alpha} 
\exp \left\{ -\varepsilon_{\alpha}/\Lambda \right\},
\eeq
where $\eta$ is the strength of the coupling to the bath, and
$\Lambda$ is the energy cutoff for the bath degrees of freedom.
The inverse cutoff $1/\Lambda$ represents the time scale for
inertial effects in the bath \cite{Weiss}.
We note that the model we consider is similar to that of trapped
impurity atoms in a Bose-Einstein condensate considered
in Ref.~\cite{Bruderer}.

We focus on the mean-field phase diagram that, 
in the absence of the bath, can be obtained by decoupling the 
tunneling term $\hat{H}_J$ and finding self-consistently the condition 
for a non-zero mean-field superfluid order parameter $\Psi_B$.  Our main goal
here is to elucidate the qualitative effects of the coupling to a bath, hence 
our use of mean field theory -- 
future work will investigate effects beyond mean field theory.
Identical results are obtained if one derives
the partition function with the effective action that describes
fluctuations of $\Psi_B$ close to the transition line in the Mott phase. 
Following the line of Refs.~\cite{Sachdev, DeMarco},
we write down the effective mean-field Hamiltonian:
\beq\label{eqnmf}
\hat{H}_{{\rm MF}}  = \hat{H}_0 + \hat{H}_{bath}+\hat{H}_{coup}
-\sum_i \left( \Psi_B \hat{a}_i^\dagger + \Psi^*_B \hat{a}_i
\right) .
\eeq
The optimum value for $\Psi_B=zJ \langle \hat{a}_i \rangle$ and is in fact
proportional to the superfluid density, with $z$ being
the coordination number. This can be seen, if we add and subtract
$\hat{H}_{{\rm MF}}$ from $\hat{H}$, so that the mean-field value 
of the ground state energy per site is \cite{Sachdev}
\beq\label{grstmf}
{\tilde E}_{0}= {\tilde E}_{{\rm MF}} (\Psi_B)
-zJ \langle \hat{a}_i^{\dagger} \rangle \langle \hat{a}_i \rangle
+\langle \hat{a}_i \rangle \Psi^*_B 
+\langle \hat{a}_i^{\dagger} \rangle \Psi_B.
\eeq
${\tilde E}_{{\rm MF}} (\Psi_B)$ is determined by the occupation number
$n_0 (\mu/U)$ in the corresponding Mott region, but here it is 
also a function of the parameters $\eta$ and $\Lambda$ characterizing the bath.
Since terms involving $\Psi_B$ and $\Psi^*_B$ can be treated as a
perturbation in the Mott-insulator phases, we can expand 
${\tilde E}_{{\rm MF}}$ in powers of $\Psi_B$ and $\Psi^*_B$ \cite{VanOosten}
\beq\label{expa}
{\tilde E}_{{\rm MF}} = {\tilde E}_{{\rm MF}}^{(0)}+
\chi \left| \Psi_B \right|^2 + {\cal O} \bigl( 
\left| \Psi_B \right|^4 \bigr).
\eeq
In general $\chi$ is a function of all parameters characterizing the local 
part of the Hamiltonian.
Self-consistently minimizing Eqs.~(\ref{grstmf}) and (\ref{expa}), we obtain
that $\Psi_B$ has a positive non-zero value once the 
condition $1/zJ=\chi$ is satisfied \cite{Sengupta}. It is not difficult 
to see from the perturbation series in Eq.~(\ref{grstmf}) that $-\chi$ 
coincides with the zero-frequency Fourier component of the 
on-site Green's function for bosons at $T=0$,
\beq\label{GF}
i {\mathcal G} (t-t^\prime) =\langle \hat{T} \hat{a}_i (t) 
\hat{a}_i^{\dagger} (t^\prime)\rangle .
\eeq
In Eq.~(\ref{GF}), $\hat{a}_i (t)$ are Heisenberg operators and
the angular brackets mean that the average is performed over the 
ground state of the system involving the bosons and bath. 

To proceed, we single out the local part of the Hamiltonian 
$\hat{H}_{l}= \hat{H}_0+\hat{H}_{bath}+\hat{H}_{coup}$ 
in Eqs.~(\ref{hamil1})-(\ref{hamilin}), and employ the well-known
Lang-Firsov transformation to obtain the new local Hamiltonian 
\beq\label{lf}
\hat{H}^{\prime}_{l}= \hat{S}\hat{H}_{l}\hat{S}^{-1}, \quad
\hat{S} =\exp \left\{ \sum_{i\alpha} \frac{g_{\alpha}}{\varepsilon_{\alpha}}
 \hat{n}_i ( \hat{b}_{i\alpha}^\dagger - \hat{b}_{i\alpha}) \right\}.
\eeq
After standard manipulations \cite{Lang}, we obtain
\begin{eqnarray}
\hat{H}^{\prime}_{l} & = &
\sum_i \frac{U^{\prime}}{2} \hat{n}_i \left( \hat{n}_i -1\right)
-\mu^{\prime} \sum_i \hat{n}_i + 
\sum_{i\alpha} \varepsilon_\alpha \hat{b}_{i\alpha}^\dagger 
\hat{b}_{i\alpha} , \nonumber \\  & & \label{newloc}
\end{eqnarray}
where
\beq\label{reno}
U^{\prime} =U -2 \sum_{\alpha} \frac{g_{\alpha}^2}{\varepsilon_{\alpha}},
\quad \mu^{\prime} =\mu + 
\sum_{\alpha} \frac{g_{\alpha}^2}{\varepsilon_{\alpha}}.
\eeq
We assume weak coupling to the bath which implies that $U^{\prime}$ is 
positive.
The transformation Eq.~(\ref{lf}) means that 
the Green's function Eq.~(\ref{GF}) can be re-expressed in 
terms of the operators
%the new dressed operators
$\hat{a}_{i}^{\prime}=\hat{S}\hat{a}_i \hat{S}^{-1}$, 
defined in the Heisenberg representation with respect 
to $\hat{H}^{\prime}_{l}$. Subsequent calculations are 
tedious but similar to those considered in 
Ref.~\cite{Mahan}. 
Hence,
\beq
i {\mathcal G} (t-t^\prime) = i {\mathcal G}_0 (t-t^\prime)
F (t-t^\prime),
\eeq
with
\begin{eqnarray}
F (t-t^\prime) & = & \exp \left\{ 
-\sum_{\alpha} \left( g_{\alpha}/ \varepsilon_{\alpha} \right)^2
\left( 1- e^{-i\varepsilon_{\alpha} |t-t^\prime|} \right) \right\}, \nonumber \\
& & 
\end{eqnarray}
and
\beq
i {\mathcal G}_0 (t-t^\prime)& = &
(n_0 +1) \,  e^{i |\xi_{-}|(t^\prime-t) } \theta (t-t^\prime)\nonumber\\
&  & + n_0 \,  e^{i \xi_{+}(t-t^\prime) } \theta (t^\prime-t),
\eeq
where $\xi_\pm$ are the energies of particle-hole excitations
\beq\label{xis}
\xi_{+}=\mu^\prime -U^\prime (n_0 -1)>0, \quad 
\xi_{-}=\mu^\prime -U^\prime n_0<0.
\eeq
At $T=0$, the ground state occupation for bosons in Mott-insulating
phases is determined as
$n_0=n_0 (\mu^\prime /U^\prime)= {\rm Integer}[\mu^\prime/U^\prime]+1$, or
$n_0=0$, if $\mu^\prime<0$, where $U^\prime$ and $\mu^\prime$ are defined in
Eq.~(\ref{reno}). 
Using Eq.~(\ref{inter}), one can easily show that 
$(\sum_{\alpha} \rightarrow \int_0^{\infty} d\varepsilon )$
\beq
F (t-t^\prime) = \left( 1+i\Lambda |t-t^\prime| \right)^{-\eta},
\eeq
so that the equation for the phase boundary 
separating the insulating and superfluid 
phases is recast as
\begin{eqnarray}
\frac{1}{zJ} & = & -i{\mathcal G} (\omega = 0) \nonumber \\
& = & \int_0^\infty \frac{dx}{(1+x\Lambda)^\eta} \left[
(n_0 +1) e^{-x|\xi_{-}|} +n_0 e^{-x\xi_{+}} \right], \nonumber \\
& &
\label{phb}
\end{eqnarray}
where $x$ is an auxiliary variable of integration. The integral in
Eq. (\ref{phb}) can be further expressed as a linear combination 
of confluent hypergeometric functions, albeit we have found
it more convenient to integrate numerically to obtain the $T=0$ 
phase diagram.

The form of the coupling in Eq.~(\ref{inter}) implies that
$U^\prime =U -2\eta \Lambda$ and $\mu^\prime =\mu +\eta \Lambda $. This leads
to two new important parameters $\eta$ and $\eta \Lambda /U$ in the 
problem as a result of coupling to a bath. We only
 consider the situation in which $\eta \Lambda /U \ll 1$,
so that the suppression of the bare interaction $U$ is not too strong. This 
means that $\eta$ can be comparable to unity only if
$\Lambda /U \ll 1$. Such a case corresponds  physically to a narrow 
band of low-energy bath degrees of freedom strongly interacting with
the lattice bosons, and is different from the picture
in which $\Lambda  \sim O(U)$ but the coupling is weak, $\eta \ll 1$.
%This latter scenario is more likely to be experimentally relevant, although
The first scenario is close in spirit to the model considered in Ref.~\cite{Bruderer},
although in both cases novel features appear on the mean-field phase diagram.

 \begin{figure}
\includegraphics[width=6cm, height=8cm, angle=-90]{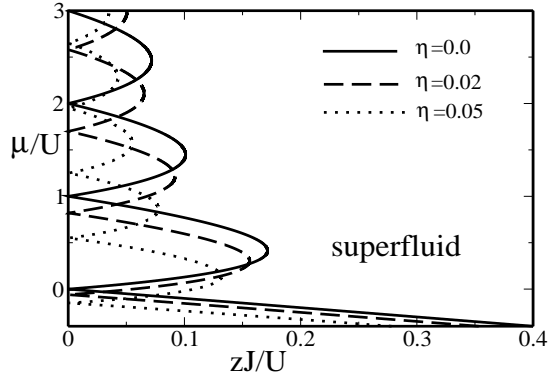}
\caption{Phase diagrams in the $zJ/U$-$\mu/U$ plane for $\Lambda/U=3.0$, and 
$\eta =$ 0.0, 0.02, and  0.05. 
The internal parts of all lobes correspond to Mott phases with
occupation number $n_0= {\rm Integer}[\mu^\prime /U^\prime]+1$.}
\label{fg1}
\end{figure}
\begin{figure}
\includegraphics[width=6cm, height=8cm, angle=-90]{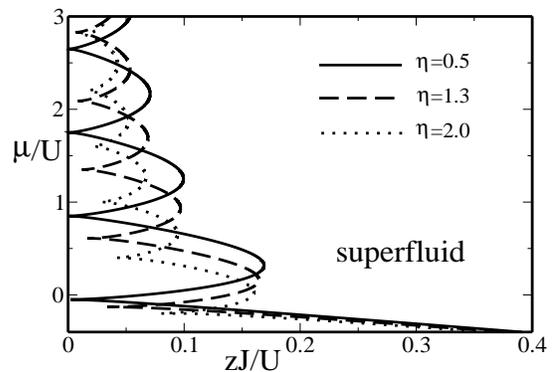}
\caption{Phase diagrams for $\Lambda/U=0.1$ and 
$\eta =$  0.5, 1.3, and 2.0. 
The meaning of regions is as in the previous figure.
For $\eta =$ 1.3 and 2.0, the horizontal lines between $zJ=0$ and 
$zJ_{-}$ separating the Mott regions with $n_0$ differing by one
 are not shown.}
\label{fg2}
\end{figure}
Figures~\ref{fg1} and~\ref{fg2} display 
mean field phase diagrams calculated from Eq.~(\ref{phb}) in the
limits of $\Lambda/U > 1$ and $\Lambda/U \ll 1$ respectively, for
several values of $\eta$, such that $\eta\Lambda/U < 0.25$ in all cases.
Since $n_0$ is determined by $\mu^\prime$ and $U^\prime$, a given occupation
number corresponds to lower $\mu/U$ in comparison to the case in which
$\eta =0$. 

An important feature to note is the  difference in the 
functional behavior of the phase boundary near the points of integer 
$\mu^\prime/U^\prime$ for a) weak bath coupling, $\eta <1$; and b)
strong bath coupling, $\eta >1$. 
From Eq.~(\ref{phb}) it follows that for $\eta >1$, the integral over $x$ 
converges even if we put $\xi_{+}$ or $\xi_{-}$ equal to zero, in contrast 
with the case $\eta <1$. We consider these two cases separately and find 
the corresponding behavior close to the degeneracy points. 
If one moves along the upper branch of the lobe corresponding to the 
occupation number $n_0$ towards the point  $(\mu^\prime/U^\prime)=n_0$ from
below,
$|\xi_{-}|\rightarrow 0$ and $\xi_{+}\rightarrow U$. At the same time,
approaching the same point from above following the lower branch of the
$n_0 +1$ lobe, $|\xi_{-}|\rightarrow U$ whilst $\xi_{+}\rightarrow 0$.  
Hence: a) If $\eta <1$, it is easy to find the corresponding asymptotics
for $zJ$ from Eq. (\ref{phb}) for small $|\xi_{-}|$ or $\xi_{+}$: close to the
point $(\mu^\prime/U^\prime)=n_0$, $zJ \propto |\xi_{-}|^{1-\eta}$ for the 
lower branch, and  $zJ \propto \xi_{+}^{1-\eta}$ for the upper one.
This behavior suggests a slightly narrower region of 
superfluid phase between the adjacent Mott lobes 
than would occur for $\eta =0$. 
b) For $\eta >1$, there is never a superfluid at $zJ=0$
and a finite value of $zJ$ is required to achieve superfluidity for any
value of the chemical potential $\mu$.
Denoting as $zJ_{-}$ and $zJ_{+}$ the limits obtained exactly at 
$(\mu^\prime/U^\prime)=n_0$ for the lower and upper branches respectively,
we find from Eq.~(\ref{phb}), that
\beq
\frac{1}{zJ_{\pm}} =\int_0^\infty \frac{dx}{(1+x\Lambda)^\eta} \left[
(n_0 +1)+ (n_0+1 \pm 1) e^{-xU} \right],\nonumber
\eeq
These two limiting values are not equal 
to each other: $zJ_{+} > zJ_{-}$,
and for $\Lambda/U \ll 1$ differ from each other 
by $O (\Lambda^2 /U^2)$. At the lowest order in $\Lambda/U$, % though
$zJ_{\pm} \approx (n_0 +1)(\eta -1)(\Lambda/U)$.
The horizontal line between $zJ_{+}$ and $zJ_{-}$, albeit short, 
separates the superfluid phase from the Mott insulator having occupation
number equal to $n_0$. 
There is also a line between $zJ =0$ and 
$zJ_{-}$ (not shown in Fig.~\ref{fg2}) which separates Mott states
with occupation $n_0$ and $n_0 +1$.
These two features of the mean-field diagram are new and arise as 
a result of strong interaction between the bosons and bath. 

It should be emphasised that whilst our results are mean field
expressions for the BHM coupled to a bath at zero temperature, they
form the basis for further investigations of effects beyond mean field
theory and at finite temperatures.
We expect that the addition of fluctuations beyond mean field theory
will modify our expressions for the phase boundary between Mott insulating
and superfluid phases, although the result that a finite
value of $J$ is required for superfluidity at all $\mu$ will be robust.
At low, but finite temperatures, there will be
depletion of the condensate and, as a consequence, enlargement 
of non-superfluid parts of the phase diagram. Studies of the BHM 
(in the absence of a bath) indicate that the changes from the $T=0$ 
case are largest around the points where $\mu/U$ is an 
integer \cite{Gerbier1}.  Hence, in analogy with Ref.~\cite{Gerbier1}
we expect that, for $\eta <1$, superfluidity will be suppressed around
the points $(\mu^\prime/U^\prime)=n_0$. For $\eta >1$, in its turn,
the transition between $n_0$ and $n_0+1$ Mott states at small $J$ becomes a 
crossover, while the superfluid regions on the whole will be 
shifted towards larger $zJ/U$.  

In addition to the general features we have outlined above, it is also
of interest to connect our results to experiments in optical lattices.
These experiments are performed in the presence of confining magnetic traps, 
rather than the uniform system considered above. This means 
that the trap potential $-V(\vec r)$ must be added to the
chemical potential $\mu$. This leads to the appearance of the 
so-called ``wedding cake'' structure \cite{DeMarco,Folling} consisting of 
concentric regions of Mott phases separated by shells of 
superfluidity in which atoms are mobile. Our results indicate that coupling 
to the bath decreases the thickness of the shells, thus facilitating the 
crossover from superfluid to insulator upon increasing $U/J$. 

It should be emphasized that the model discussed here
is independent of other mechanisms of dissipation, such as 
excitation of the phase slips in a moving condensate observed 
recently \cite{McKay,Polkovnikov}. If a heat bath plays a
role in experiments on cold atoms in optical lattices, then it would
be highly desirable to have a way to determine its strength.
 Unfortunately, however,
for the model considered here, it is difficult to 
determine from the currently available experimental data how strong 
interaction with the bath may be. However, we believe  
that this particle-conserving interaction is not unrealistic 
on timescales relevant to experiments, and might be detected
by performing lattice depth modulation experiments similar 
to those discussed in Ref.~\cite{Schori}. 
In these experiments, the lattice potential is modulated at 
a fixed frequency 
for a short period of time. The modulation effectively translates into 
variation of $J$. During this process, energy is transferred to the 
atoms resulting in heating and the broadening of the 
central momentum peak, indicating an
increase in the non-superfluid fraction of the atoms as observed after
ballistic expansion once the trap is switched off.
The presence of coupling to a bath
 may mean that a system heated quickly by modulation 
will cool with time. Hence, varying the time between the
end of the modulation and the free expansion
may reveal a decrease of the out-of-condensate portion 
as a result of relaxation towards the initial equilibrium state 
giving a window to access the bath coupling.

In conclusion, we have performed an analysis of the effects of
coupling the BHM to an ohmic bath on the mean field phase diagram
of the BHM.  We find that the changes from the absence of a bath
depend on both the strength of the coupling to the bath and the 
width of the bath spectrum.  We believe our work should be 
relevant to any physical system that can be modelled with a BHM.

We thank C. Chamon, J. McGuirk, and Kun Yang for discussions 
relating to choice of coupling to the bath and dissipation in 
cold atom systems. This work was supported by NSERC.

\end{document}